\documentclass[onecolumn]{emulateapj}
\usepackage{natbib}
\usepackage{amsmath,amsfonts}
\usepackage{epsfig}
\usepackage{multirow}
\usepackage{mathptmx}
\usepackage{longtable}

\begin{document}

\preprint{APS/123-QED}

\title{Radio Emission from Low Mass Young Stellar Objects}
\author{Anna M. M. Scaife}
\affiliation{Physics \& Astronomy, University of Southampton, Highfield, Southampton, SO17 1BJ, UK\\ email:~a.scaife@soton.ac.uk}

\date{\today}

\begin{abstract}
Compact radio emission provides a reliable method for the detection of low luminosity young stellar objects (YSOs), and is particularly useful for detecting the earliest stages of protostellar evolution where the source itself may still be heavily embedded in its natal dust envelope. For such Class~0 and Class ~I objects the dominant radio emission mechanism is expected to be free-free, however unlike massive YSOs the way in which this radio emission is produced remains a subject of debate. As larger samples of radio YSOs become available the relationship between the radio luminosity of the Class~0/I population and their wider global properties is now being clarified. Furthermore, the broader scientific applications of such samples are also becoming increasingly apparent. These improved constraints on the nature of the radio emission from YSOs are now contributing to our understanding of not
only the evolutionary physics of protostars themselves but also their wider impact on their surroundings. Here we discuss the physics of the radio emission, the emerging relationship between this emission and other properties of YSOs and some of the applications for studies exploiting this emission.
\end{abstract}


\maketitle

\section{Introduction}
\label{sec:rad}

The long wavelength radio emission from low-mass young stellar objects (YSOs; Class 0 \& Class I) is capable of escaping from the high column density dust envelopes which surround these deeply embedded young protostars, making such objects detectable in the radio band even when opacity effects conceal them at infrared frequencies. Although the thermal dust spectrum falls off steeply in intensity at longer wavelengths, young and low-luminosity protostars are known to produce additional radio emission. This radio emission has been observed to possess a rising spectrum, indicating that it occurs as a consequence of free-free radiation from an ionized plasma. Although the physics behind the ionization around high-mass stars is reasonably well understood as arising from photoionization due to the strong UV flux from such objects, in the low-mass case a number of mechanisms have been proposed to produce the ionization responsible for the observed free-free emission. Distinguishing between these possible mechanisms is not trivial from currently available data. Where a high enough ionizing flux is present, generally in later type T~Tauri stars, photoionization can also support an embedded {\sc Hii} region (Churchwell 1990); a fully ionized stellar wind, associated more often with more evolved stars than very young YSOs, would produce an observable radio signal (Panagia \& Felli 1975; Curiel et~al. 1989), as could a partially or fully ionized, collimated outflow (Reynolds 1986); the accretion shock on the surface of protostellar discs may heat and ionize infalling gas (Winkler \& Newman 1980; Cassen \& Moosman 1981); molecular outflows from young, low-mass protostars and the neutral winds thought to initiate them can generate free-free emission through shock ionization as they impact on the surrounding envelope (Curiel, Canto \& Rodr{\'i}guez 1987; 1989; Rodr{\'i}guez \& Reipurth 1996), a mechanism which was refined by Ghavamian \& Hartigan (1998).

It is often the last of these mechanisms which is considered to be most plausible and indeed, where measured, the outflow force of many protostellar jets has been found to be energetically viable to explain the observed cm-wave radio emission (Anglada 1995). This theory is also supported by very high angular resolution observations of the radio emission from protostars (Simon et al. 1983; Anglada 1996), which often find that the radio emission is elongated along the direction of the molecular outflows. However, as increasingly sensitive molecular observations reveal weaker and weaker outflows (e.g. Bourke et~al. 2005; Crapsi et~al. 2005) the measured momentum flux is also found to be insufficient to explain the total radio emission in a rising number of objects (e.g. Scaife et~al. 2011a) and it is possible that a combination of the above mechanisms are infact responsible for the observed radio emission.

As mentioned above, it is generally assumed that the dominant mechanism giving rise to continuum radio emission from YSOs is thermal bremsstrahlung, or \emph{free-free}, emission. This mechanism operates when charged particles, typically electrons, are accelerated by encountering another charged particle. Consequently radio free-free emission is expected to manifest from any environment populated by an ionized plasma. The distribution and energy cut-off for scattering in such a plasma is characterized by the Gaunt factor, $\langle g_{\rm ff} \rangle$, which allows one to work out the (ratio of) absorption coefficient, $\kappa_{\nu}$, for the plasma.Integration of this quantity along the line of sight, $\int{\kappa_{\nu} {\rm d}\ell}$, gives the optical path length (depth) for the free-free emission, $\tau_{\nu}$. Optical depth is a strong function of electron density, generally parameterized through the emission measure of the plasma, $EM = \int{N_{\rm e} N_{\rm i} {\rm d}\ell }$, and a weaker function of electron temperature. Once the optical depth of the emitting plasma has been determined the brightness temperature, $T_{\rm b}$, of the resulting emission is given by the product,
\begin{equation}
\label{equ:tb}
 T_{\rm b} = T_{\rm e}(1-{\rm e}^{-\tau_{\nu}}).
\end{equation}
In the Rayleigh-Jeans region of the spectrum the resulting flux density is then given by $S_{\nu} = 2k_{\rm B}T_{\rm b}/\lambda^2 \Omega$.

Where the plasma giving rise to the free-free emission is reasonably uniform in density this leads to a characteristic radio spectrum which has two components delineated by the frequency at which the optical depth equals unity ($\tau_{\nu} = 1$), marking the transition from optically thick behaviour to optically thin. In the optically thick regime the optical depth term becomes approximately unity and the flux density spectrum rises as $\nu^2$; in the optically thin regime the optical depth term becomes $\approx \tau_{\nu}$ and the frequency dependence of $\tau_{\nu}$ cancels with the explicit frequency dependence in the Raylegh-Jeans approximation, resulting in a flux density spectrum which varies as $\nu^{-0.1}$.  

Key to free-free emission is the presence of a hot, ionized plasma. For massive YSOs photoionization of surrounding neutral gas by the ionizing UV flux from the central object is high enough to account for the observed radio emission. For low mass, low luminosity objects this is not the case. Fig.~\ref{fig:lbol} shows the expected radio luminosity from the Lyman continuum flux of a given bolometric luminosity `zero-age-main-sequence' (ZAMS) star. This comparison was first shown by Anglada (1995) and following that work we adopt Lyman continuum fluxes from  Thompson (1984) for a given range of bolometric luminosities. The radio luminosity at a given frequency can then be derived using
\begin{equation}
\left(\frac{L_{\rm rad}}{{\rm mJy\,kpc^2}}\right) = 2.08\times 10^{-46} \left(\frac{N_{\rm Lyc}}{\rm cts\,s^{-1}}\right)\left(\frac{\nu}{\rm GHz}\right)^{-0.1}\left(\frac{T_{\rm e}}{\rm K}\right)^{0.45},
\end{equation}
which has been adapted from Wilcots (1994). Here we assume $T_{\rm e}=10^4$\,K and update the measured data to include objects at lower luminosities from the AMI-LA radio continuum follow-up (AMI Consortium: Scaife et~al. 2011 a;b; 2012) to the \emph{Spitzer} c2d survey (Evans et~al. 2005).

It is immediately obvious from Fig.~\ref{fig:lbol} that the radio luminosity of low bolometric luminosity YSOs is significantly in excess of that expected from photoionization, and that this trend continues to very low ($L_{\rm bol} \ll 1$\,L$_{\odot}$) luminosities where the discrepancy reaches $>10$ orders of magnitude.

The alternative explanations for the radio emission associated with these low luminosity YSOs as listed above generally invoke the action of powerful molecular outflows and stellar winds associated with such objects. The presence of these energetic processes provides an alternate source for the ionization required to explain the observed free-free emission. 

In these cases the ionized plasma generally cannot be well approximated by uniform optical depth but instead has regions of varying density, resulting in a \emph{partially} opaque plasma. The spectra from such regions have behaviour that depends not only on the thermodynamic conditions, but also on the geometry of the ionized gas. 

The canonical geometry for a stellar wind is that of a spherical region of ionized, isothermal\footnote{The polytropic case for the same geometry was considered by Chiuderi \& Torricelli~Ciamponi (1978), whose derivation pre-empted the analysis of Reynolds (1986) but in a less general form.} plasma with a radial density gradient, $n_{\rm e}(r) = n_0 (r/r_0)^{-2}$ (Panagia \& Felli 1975), where $n_0$ is the electron density at the inner boundary $r_0$, which marks the stellar radius. For such an emission region the optical depth through the plasma varies as a function of projected radius on the sky.

For a generalised outflow where the physical conditions are parameterized by different radial dependencies such that $T_{\rm e}(r) = (r/r_0)^{q_T}$, $\tau(r) = (r/r_0)^{q_{\tau}}$ etc the spectral index of the radio emission would have the form
\begin{equation}
\alpha = 2 + \frac{2.1}{q_{\tau}}(1+\epsilon+q_T)
\end{equation}
with $q_{\tau} = \epsilon + 2q_x + 2q_n - 1.35q_T$ (Reynolds 1986), where $\tau$ denotes optical depth, $x$ denotes ionization, $n$ denotes density, $T_{\rm e}$ denotes temperature and $\epsilon$ relates to the opening angle of the jet such that $w = w_0(r/r_0)^{\epsilon}$ with $w$ representing the radius of the jet perpendicular to the direction of outflow. For a range of physical situations this index varies with values $-0.1 \leq \alpha \leq 1.1$, as listed in Table~1 of Reynolds (1987). 

Opening angles for outflows from young stellar objects (YSOs) vary with evolution. For example the earliest stage of YSO (Class~0) have highly collimated outflows which become less collimated as they evolve through Class~I and, where an outflow remains, leave a wide angle outflow from Class~II objects. Typically, however, one may assume that $\theta_0 \leq 0.5$\,rad. Outflows which are collimated ($\epsilon <1$) will have shallower radio spectral indices, typically $\alpha=0.25$, although pressure confined outflows can exhibit much steeper indices approaching unity. Such steep indices can also be produced by recombining or accelerating outflows where the indices may even exceed unity. A conical, fully ionized, isothermal wind will have the same spectral index as the canonical spherical stellar wind of Panagia \& Felli (1975): $\alpha=0.6$. This is assumed to be the typical spectral index for radio emission from YSOs, and indeed the radio spectral indices from samples of YSOs, where measurable, typically have a mean value of 0.6 (e.g. Scaife et~al. 2012).

\section{Characteristic Correlations}
\label{sec:corr}

\begin{figure}
\centerline{\includegraphics[width=0.5\textwidth]{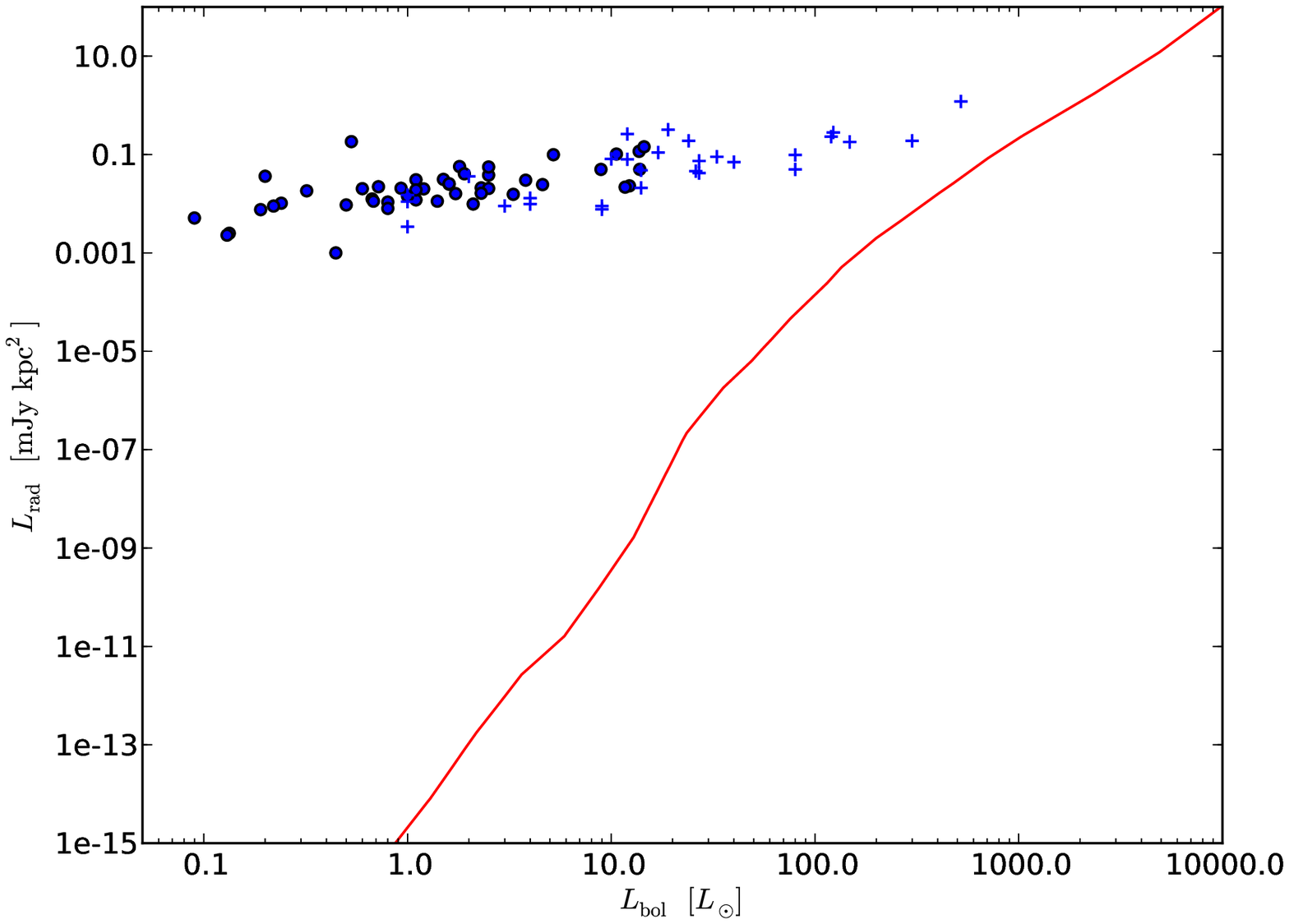} \qquad \includegraphics[width=0.5\textwidth]{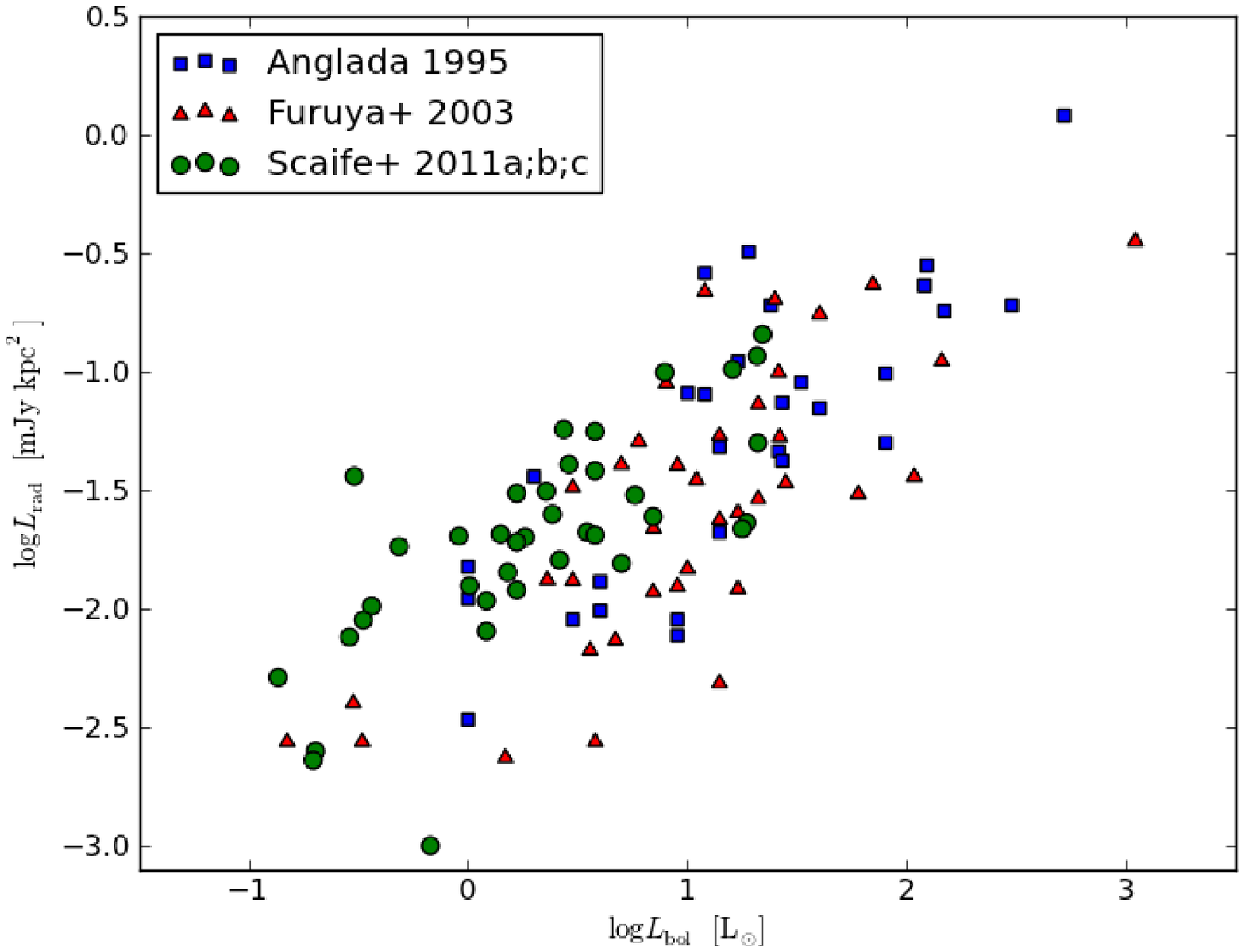}}
\centerline{(a) \hskip 0.5\textwidth (b) }
\caption{Observed radio luminosity as a function of bolometric luminosity. (a) The solid line shows the radio emission expected from Lyman continuum radiation from a ZAMS star, see text for details. Data points are from the AMI-LA (filled circles; AMI Consortium: Scaife et~al. 2011a; b; 2012) and from the compilation of Anglada (1995; crosses). (b) $L_{\rm rad}-L_{\rm bol}$ correlation showing compiled data from the literature normalized to 5\,GHz assuming a global spectral index of $\alpha=-0.6$. \label{fig:lbol}}
\end{figure}

In most cases the number of free parameters in detailed theoretical models of partially optically thin free-free emission far exceed the physical constraints available from observational data. Consequently it is typical to express the expected radio luminosity as a function of alternate observables. 

It can be seen from Fig.~\ref{fig:lbol} that there is a strong correlation between radio and bolometric luminosity ($r=0.77$; Scaife et~al. 2011a), characterized as 
\begin{equation}
\label{eq:lrad}
\log L_{\rm rad} [{\rm mJy\,\,kpc^2}] = -(1.74\pm0.18)+(0.51\pm0.26)\log L_{\rm bol} [{\rm L_{\odot}}],
\end{equation}
at a frequency of 16\,GHz (Scaife et~al. 2011b). An even stronger correlation ($r=0.89$; Scaife et~al. 2011b) is evident when considering the relationship between radio luminosity and envelope mass,
\begin{equation}
\log L_{\rm rad} [{\rm mJy\,\,kpc^2}] = -(2.23\pm0.65)+(0.68\pm0.62)\log M_{\rm env} [{\rm M_{\odot}}]
\end{equation} 
(Scaife et~al. 2011b). Similarly to the correlations of outflow force, $F_{\rm CO}$, with bolometric luminosity and envelope mass characterized by Bontemps et~al. (1996) there is a stronger dependency on evolutionary state indicated in the relationship between radio luminosity and envelope mass, compared to that with bolometric luminosity. The power-law index of the relationship $L_{\rm rad} \propto M_{\rm env}^{0.7}$ is dominated by the presence of Class~I objects, which deviate from the general trend (Scaife et~al. 2011b). When considering Class~0 objects alone the power-law index is significantly increased and in both cases is in fact consistent with that expected from models of both spherical (Panagia \& Felli 1975) and collimated (Reynolds 1986) stellar winds, which predict that
\begin{equation}
L_{\rm rad} \propto M_{\rm env}^{4/3},
\end{equation}
as for simple models of protostellar evolution the envelope mass is expected to be proportional to the mass loss rate (Bontemps et~al. 1996; Fuller \& Ladd 2002; Arce \& Sargent 2006).

Anglada (1995) suggested that there should also be a correlation of radio luminosity with outflow force, as might be expected from the shock ionization models of Curiel et~al (1989). Where molecular outflows from young, low-mass protostars and the neutral winds thought to initiate them generate free-free emission through shock ionization as they impact on the surrounding envelope (Curiel, Canto \& Rodr{\'i}guez 1987; 1989; Rodríguez \& Reipurth 1996) the free-free flux density would be proportional to the outflow force (also called momentum flux), $F_{\rm CO}$, which is equivalent to the rate of outflow momentum and is often calculated as $F_{\rm CO}= p_{\rm CO}/\tau_{\rm dyn}$. 

At a wavelength of 2\,cm the expected flux density from such shock ionization can be computed as
\begin{equation}
\left(\frac{S_{\rm \lambda=2\,cm}}{\rm mJy}\right) = 3.4\times 10^3 \eta \xi(\tau)^{-1}\left(\frac{F_{\rm{CO}}}{\rm{M}_{\odot}\,\rm{yr}^{-1}\,\rm{km\,s}^{-1}}\right)\left(\frac{d}{\rm kpc}\right)^{-2},
\end{equation}
(Scaife et~al. 2010; Anglada 1996), where $\eta=(\Omega/4\pi)$ is the fraction of the stellar wind being shocked and the factor $\xi(\tau)=\tau/(1-\rm{e}^{-\tau})$ allows for the fact that the radio emission may not be optically thin (Anglada et~al. 1998), although this dependence on the optical depth, $\tau$, is very weak. 

\begin{figure}
\centerline{\includegraphics[width=0.5\textwidth]{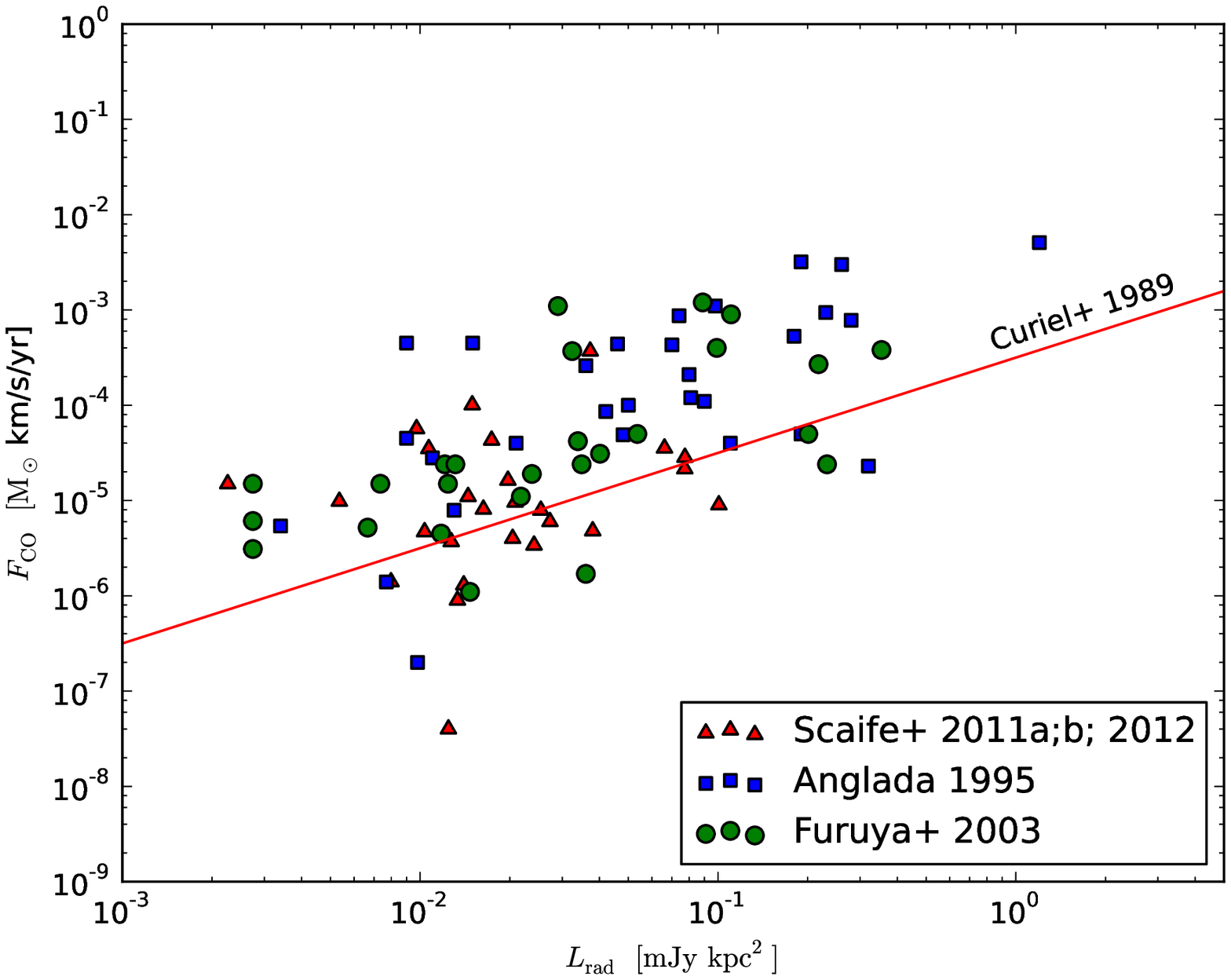}\qquad\includegraphics[width=0.5\textwidth]{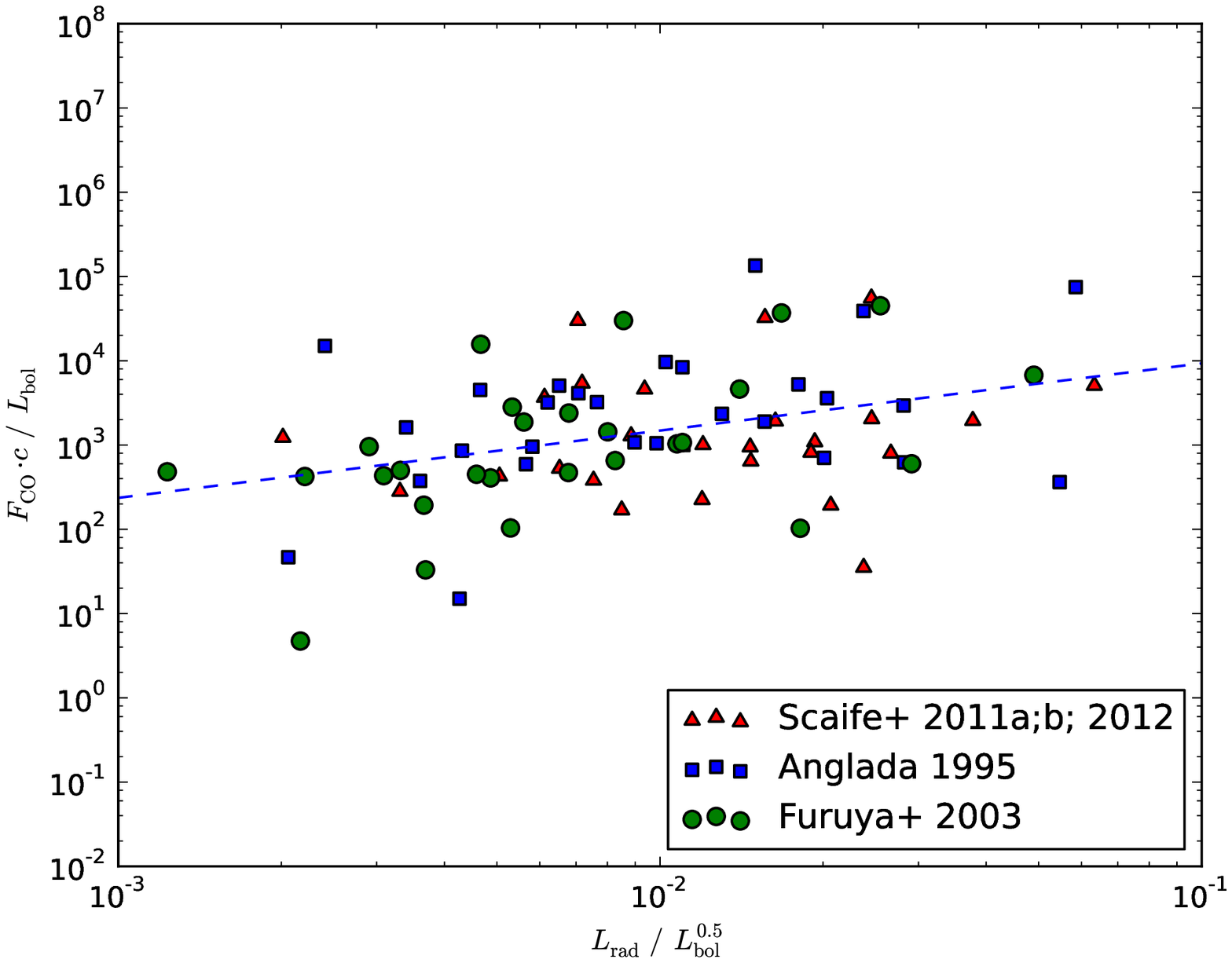}}
\centerline{(a) \hskip 0.5\textwidth (b)}
\caption{Observed radio luminosity as a function of outflow force. (a) Compiled data from the literature are shown normalized to 8.35\,GHz assuming a global spectral index of $\alpha=-0.6$. The minimum outflow force required to produce a given radio luminosity is indicated for plasma with electron temperatures of $3000$\,K (dashed line) and $10^4$\,K (solid line). (b) $F_{\rm CO}c/L_{\rm bol}$ (dimensionless) versus $M_{\rm env}/L_{\rm bol}^{0.6}$ for all objects in the listed data sets that have available values for all three physical parameters. \label{fig:fcorr}}
\end{figure}

Fig.~\ref{fig:fcorr}(a) shows the distribution of observed radio luminosity with outflow force for data compiled from the literature. A weak positive correlation is present in these data with $r=0.62$ and a best fitting regression of $F_{\rm CO} \propto L_{\rm rad}^{1.09\pm0.15}$, consistent with that determined by Anglada (1995) for the original dataset. Removing the luminosity dependence from these data by using the relationship determined by Bontemps et~al (1996), $F_{\rm CO} \propto L_{\rm bol}$, and that shown in Eq.~\ref{eq:lrad}, appears by eye to improve the correlation, see Fig.~\ref{fig:fcorr}(b) but in fact results in an even poorer Pearson co-efficient ($r=0.36$). In both cases the regression is consistent with that expected for the shock ionization model, but the correlation of radio luminosity with outflow force remains very weak with substantial scatter. 

There are a number of effects, resulting from the different methods used to calculate outflow force, that will add scatter to these distributions. While some estimates are derived from the total momentum observable in the outflow lobes, divided by a measure of the outflow dynamical time, others use a measure of the force derived within an annulus around the outflow origin (Bontemps et~al. 1996). This later provides a better estimate of the force for outflows which have been only partially mapped. Further variations arise from whether the maximum velocity in an outflow is used to compute the dynamical time or some average/typical velocity. Flux loss in interferometric studies, high optical depths of CO emission in the outflow line wings and inclination effects will all also reduce the observed outflow velocities and tend to reduce the momentum actually observed. Many studies follow the lead of Bontemps et~al. (1996) and increase their outflow forces by a factor of 10 to account for optical depth and inclination effects (e.g. Visser et~al. 2002). Optical depth effects can also be corrected for using observations of the rarer isotopologues of CO (e.g. L1221-IRS3; Umemoto et al. 1991), and some studies note that the inclination corrections will be minimal (e.g. L1014-IRS; Bourke et~al. 2005). 

In addition to the lack of a strong correlation between radio luminosity and outflow force, a further concern from these data is the number of objects which appear to have outflow force values below the minimum level required to explain their radio luminosity via the shock ionization model. This level is indicated by a solid line in Fig.~\ref{fig:fcorr}(a). Such sources are evident even in the initial compilation of Anglada (1995) and the population has become more pronounced as an increasing number of objects with weaker outflows have been detected (Scaife et~al. 2011b; 2012). This discrepancy is important as it may indicate that the shock ionization model for radio emission from YSOs is not the dominant mechanism operating in these objects. 

However, the observed discrepancy may also be an effect of the method used for calculating outflow force: it was noted by Curtis et~al. (2010) that the dynamical time may not be suitable as a divisor for calculating the outflow force. Dynamical time, defined as $\tau_{\rm d} = L_{\rm lobe}/v_{\rm max}$, is the time required for the bow shock of an outflow to travel the distance of that outflow at the maximum velocity observed in the outflow. This quantity is not a good measure of the age of the outflow. It has been suggested both to underestimate true outflow ages in evolved sources (Parker, Padman \& Scott 1991) and to overestimate outflow ages in young Class~0 sources (Downes \& Cabrit 2007). A better tracer of outflow age has been suggested to come from the bolometric temperature of a source (Curtis et~al. 2010), which is expected to increase as a protostar consumes its surrounding dust envelope and advances towards the main sequence. 

\begin{figure}
\centerline{\includegraphics[width=0.5\textwidth]{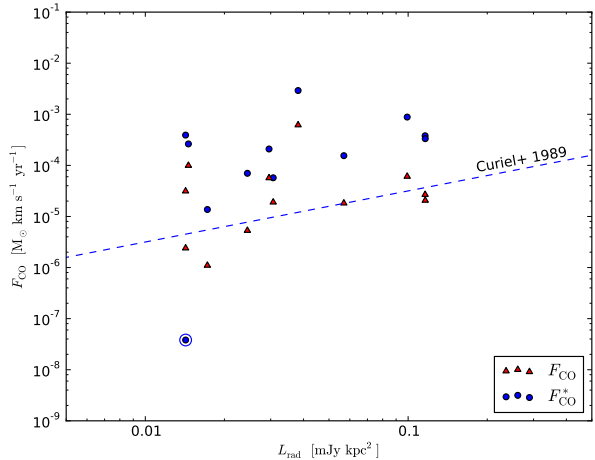}}
\caption{Data for a subsample of objects is shown with outflow forces calculated in different ways. $F_{\rm CO} = p_{\rm CO}/\tau_{\rm d}$ uses the traditional method with dynamical time as the divisor, whereas $F_{\rm CO}^{\ast} = p_{\rm CO}/t_{\rm YSO}$ uses the alternate method with the age of the driving source as the divisor, see text for details. All objects are Class~0, with the exception of the ringed source which is Class~I. The minimum outflow force required to explain the observed radio emission using the shock ionization model of Curiel et~al. (1989) is indicated by a dashed line. \label{fig:curtis}}
\end{figure}

The age of the driving source, and hence outflow age, can be estimated from the bolometric temperature using the relation,
\begin{equation}
\log t_{\rm YSO} [{\rm yr}] = (-0.9\pm0.6)+2.4\log T_{\rm bol} [{\rm K}]
\end{equation}
(Ladd, Fuller \& Deane 1998). Examining a subsample of the data presented in Fig.~\ref{fig:fcorr}(a), which has information available to calculate both the typically used outflow force using the dynamical time , $F_{\rm CO}= p_{\rm CO}/\tau_{\rm d}$, and the outflow force calculated using the outflow age, $F_{\rm CO}^{\ast} = p_{\rm CO}/t_{\rm YSO}$, it can be seen that objects which are inconsistent with the Curiel et~al. (1989) shock ionization model using the first method, become consistent when the second method is applied. This distinction is illustrated in Fig.~\ref{fig:curtis}. The one object which remains inconsistent with the shock ionization model is the only Class~I source in this sample (circled), suggesting that for Class~I objects an additional radio emission mechanism may be present. In addition, although only one Class~I object is present in this sample we note that the object currently known to be most discrepant with the shock ionization model in this regard (L1014-IRS1; Shirley et~al. 2007; Scaife et~al. 2011a), shown as the most low-lying point in Fig.~\ref{fig:fcorr}(a) but not present in the subsample, is also Class~I.

\section{Radio Selection Functions}
\label{sec:select}

\begin{figure}
\centerline{\includegraphics[width=0.5\textwidth]{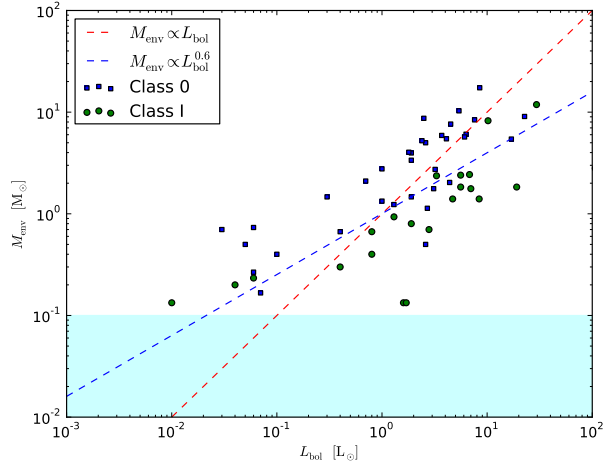}}
\caption{Relationship between envelope mass and bolometric luminosity. Data are shown for the Perseus molecular cloud from Hatchell et~al. (2007) with Class~0 objects shown in blue and Class~I objects shown in green. The envelope mass detection threshold for these data is indicated by a light blue shaded area. \label{fig:evol}}
\end{figure}

In spite of the strong correlation between radio and bolometric luminosity, this relationship is defined by protostellar objects which are detected in the radio. The distribution of \emph{non-detections} of protostellar objects does not seem to be a function of the radio luminosity detection limit when considered against bolometric luminosity, but is instead better traced by the radio luminosity detection limit considered against envelope mass (Scaife et~al. 2012). Indeed those objects which are not detected in the radio are sources that have a high ratio of bolometric luminosity to envelope mass. Such a ratio is commonly used as an evolutionary tracer amongst YSOs; in Fig.~\ref{fig:evol} we plot bolometric luminosity against envelope mass for a sample of objects from the Perseus molecular cloud (Hatchell et~al. 2007), the delineating nature of $M_{\rm env}/L_{\rm bol}$ is shown as a red dashed line and the ratio determined by Bontemps et~al. (1996) of $M_{\rm env}/L_{\rm bol}^{0.6}$ as a blue dashed line. An evolutionary distinction in radio detection rates with a ratio of $\sim M_{\rm env}/L_{\rm bol}$ is consistent with the overall radio detection rates for Class~0 and Class~I objects, which are 75 and 25 per cent, respectively (Scaife et~al. 2012). These detection ratios are common across multiple different star formation regions and, considered in light of the mass-luminosity ratio trend described previously, strongly suggest that radio detections are biased towards younger embedded objects. A consequence of this is the implication that blind radio surveys of star forming regions will preferentially detect those objects which have high ratios of envelope mass to bolometric luminosity. This makes radio surveys of such regions highly complementary to surveys in the (mid-)infared, as those data will be biased against such objects due to opacity effects. However, the very low detection rate for Class~I objects, combined with the fact that Class~I objects exhibit a degree of variability in their radio emission which is likely to arise from alternative emission mechanisms, makes this evolutionary behaviour difficult to quantitatively demonstrate at present due to small sample sizes.

\section{Radio Emission and the Evolutionary Sequence}
\label{Sec:evol}

The population of protostellar objects discussed here encompasses the earliest stages of their evolution: Class~0 and Class~I. More evolved pre-main sequence (PMS) stars, Class~II/III, are also known to exhibit radio emission (see e.g. Feigelson \& Montmerle 1999). For these later objects the radio emission is dominated by non-thermal processes associated with magnetospheric/coronal activity and the radio emission is known to be strongly correlated with X-ray emission (G{\"u}del \& Benz 1993). This correlation is thought to be due to a common underlying energy reservoir being responsible for both plasma heating that produces thermal X-ray emission and particle acceleration in the presence of magnetic fields, which prodices non-thermal radio emission. Establishing whether such a radio/X-ray relation holds for younger protostellar objects is complicated by a number of factors. Two factors which cause immediate issues are, firstly, that the thermal emission, which is observed to dominate the radio spectrum for Class~0/I objects, quickly becomes opaque to non-thermal emission (Andr{\'e} 1987) and secondly, that the high degree of variability in the emission from Class~I/II objects requires simultaneous X-ray and radio observations to be taken. Even taking these factors into account, a recent study of the radio/X-ray correlation for younger protostellar objects showed that the recovered data are inconsistent with the extrapolated G{\"u}del-Benz relation by more than an order of magnitude (Forbrich et~al. 2011). However, these results are not yet definitive as such studies are limited to small samples of objects, both due to observational constraints as well high non-detection rates for as yet unknown astrophysical reasons, as well as an uncertainty in the underlying mechanism of the detected radio emission. 

The differing radio detection rates for Class~0 and Class~I objects noted in \S~\ref{sec:select} may indicate a transition between underlying emission mechanisms during the Class~I stage, consistent with the expectation that some fraction of thermal emission may also be expected into the Class~II stage. This remnant of thermal activity in the case of Class~II objects is perhaps associated with ionization due to a surface shock (e.g. Cassen \& Moosman 1981) or ionized plasma in a disk wind (Pascucci et~al. 2012). A combination of mechanisms in the case of Class~I objects is also indicated by the polarization signature seen in case of L1014 (Shirley et~al. 2007) during flare activity, which can only have arisen from a non-thermal mechanism. The radio detection rates for Class~0/I objects are inverted in the X-ray, for example in the Serpens and NGC\,1333 regions no Class~0 objects were detected using \emph{Chandra}, whilst 23\% of Class~I objects were detected (Winston et~al. 2010). It is unclear at this stage whether this effect is due entirely to extinction/absorption of soft X-rays or if it represents a fundamental underlying trend. Future work in this area, examining the relationship between radio and X-ray emission, will be instrumental in distinguishing the underlying radio mechanisms and their distribution across protostellar class.

\section{Radio Emission as a Tool}
\label{sec:tool}

Historically the study of radio emission from low mass young stellar objects (YSOs; Class~0 \& Class~I) has concentrated on characterizing the radio emission itself and the underlying emission mechanism. However, recent advances in a number of other areas have shown that understanding, quantifying and predicting the radio emission from protostellar objects is increasingly important for a variety of reasons, including, but not limited to, the investigation and characterization of the anomalous microwave emission attributed to spinning dust, tackling the luminosity problem indicative of an incomplete understanding of protostellar accretion history, and exploring the long wavelength tail of the greybody distribution in order to understand dust grain evolution and planet formation.

The first of these, the observational study of the anomalous microwave emission attributed to spinning dust, has largely concentrated on arcminute scale emission since it is primarily targeted by CMB telescopes, which consider it a source of foreground contamination. Although dark and molecular clouds have been identified as the source of such anomalous emission (e.g. L1622, Cassasus et~al. 2006; L675 \& L1111, Scaife et~al. 2010) these large-scale observations have largely neglected the known small-scale radio emission detected from a significant fraction of protostars (Rodr{\'i}guez et al. 1989; Anglada 1995). As such investigations become more detailed it is necessary to quantify correctly this emission. Firstly, in order to investigate any connection between the presence of active star formation within a dark cloud and the identification of a spinning dust component. The observational uncertainties surrounding the identification of heavily embedded, low luminosity YSOs have lead to the proposition that many clouds currently considered to be starless, including a number not previously thought to even be undergoing collapse, do in fact host star formation. This possibility creates complications when investigating the correlation of those clouds hosting star formation with those which are associated with anomalous microwave emission, as highlighted by the case of L675 (Scaife et~al. 2010) and this remains an important open question. Secondly, in marginal cases, identifying radio emission from protostars internal to the cloud is important in order to avoid confusing their emission with any excess that may be attributed to spinning dust.

Most radio protostar searches and surveys (e.g. Andr{\'e}, Montmerle \& Feigelson 1987; Anglada 1995; Stamatellos et al. 2007) are conducted at 3.6 and 6\,cm wavelengths; however, the spectrum of emission due to spinning dust is thought to peak at shorter wavelengths of $1-2$\,cm (Draine \& Lazarian 1998) and this is the wavelength range where most spinning dust observations are made. Although historically the number of observations made at 2\,cm (Rodr{\'i}guez \& Canto 1983; Bieging, Cohen \& Schwartz 1984; Pravdo et al. 1985; Rodr{\'i}guez \& Reipurth 1998; Andr{\'e}, Motte \& Bacmann 1999) were limited, particularly in the case of low-luminosity protostars and Class~0 objects, a substantial increase in the number of such objects observed has recently been made through the \emph{Spitzer} follow-up surveys of the AMI telescope at 16\,GHz (Scaife et~al. 2011 a;b;2012). These surveys were initiated to quantify the radio contribution of, and correlation with, YSOs to the microwave emission of dark clouds following the detection of a new radio YSO in L675 (Scaife et~al. 2010).

The strong correlation between radio and bolometric luminosity first noted by Anglada (1995) has been shown by recent surveys to hold towards lower luminosities (Scaife et~al. 2012), see Fig.~\ref{fig:lbol}. Although historically radio observations of protostars have been skewed towards higher luminosity objects due to observational constraints, improvements in sensitivity have allowed this distribution to tend more towards that seen at infrared wavelengths and predicted by theoretical models of YSO evolution (e.g. Dunham \& Vorobyov 2012). One consequence of this result is that predicting the contribution of protostars, where identified, to the overall microwave emission of star-forming dark clouds may be done in principle from their bolometric luminosity in the infrared. A further area where such a relation may prove important is in determining the abundance of low luminosity protostars. 

Measurements of the luminosity distribution of protostars from the Spitzer Space Telescope (Dunham et~al. 2008; Evans et~al. 2009) have aggravated the `luminosity problem' (Kenyon et~al. 1990). This problem exhibits as an increasing number of protostars at lower luminosities down to an internal luminosity of $L_{\rm int}\simeq$ 0.1 L$_{\odot}$, below which the numbers start to decline. However, simple star formation models, such as the spherical accretion model of Shu (1977), predict that a low-mass source on the stellar/brown dwarf boundary ($M\simeq 0.08$\,M$_{\odot}$) should have an internal luminosity $L_{\rm int}\simeq 1.6$\,L$_{\odot}$ from accretion alone (Evans et~al. 2009). Non-steady accretion, starting in the earliest protostellar stages, is currently the best solution to this discrepancy (Kenyon \& Hartmann 1995; Young \& Evans 2005; Enoch et al. 2007). The luminosity problem is most difficult to rectify in very low luminosity objects (VeLLOs; Young et~al. 2004; Dunham et~al. 2008) with extreme luminosities $L_{\rm int}\leq 0.1$\,L$_{\odot}$. 

Recent predictions of the luminosity distribution expected from non-steady accretion scenarios (e.g. Dunham \& Vorobyov 2012) have instead predicted an over-abundance of low luminosity objects: the so-called `reverse luminosity problem'. These results emphasize the point that understanding the true accretion mechanism depends on observationally constraining the shape of the luminosity distribution at low luminosities. The existing sample of VeLLO sources, although expanded by \emph{Spitzer}, is in no way complete. VeLLOs are difficult to confirm as protostars in the infra-red due to their low luminosity and embedded nature. It is also the case that VeLLOs are often found in cores which are not only assumed to be starless, but which were also not believed to be approaching collapse (Bourke et ~al. 2006). Nevertheless, identifying a complete sample of these low luminosity embedded protostars is vital for understanding low mass star formation. Radio emission from heavily embedded objects provides a useful tool for establishing the presence of protostellar activity in such cases as it is not subject to the same degree of opacity effect as the shorter wavelength infra-red emission.

\section{Conclusions}
\label{sec:conc}

The long wavelength radio emission from YSOs provides a novel tool for confirming the population of low luminosity embedded protostars. Increased samples of objects have extended the known correlation of radio luminosity with bolometric luminosity by over an order of magnitude since it was first noted (Anglada 1995), and this correlation now includes a substantial number of the newly discovered VeLLO objects. A further correlation with envelope mass has also been identified. Potential problems from sources with small outflow force values for the generally accepted shock ionization model of free-free emission have been shown to disappear when outflow force is calculated using the age of the driving sources as opposed to the dynamical time. In the case of Class~I objects this discrepancy is more difficult to rectify and, when coupled with a lower detection rate, this suggests that although shock ionization may be the dominant emission source for Class~0 objects an alternate mechanism may contribute substantially to the emission of Class~I sources.

These improved constraints on the nature of the radio emission from YSOs are now contributing to our understanding of not only the evolutionary physics of protostars themselves but also their wider impact on their surroundings through e.g. their connection to the anomalous microwave emission attributed to spinning dust. As further improved observational samples become available these constraints can only improve and it is doubtless that even more applications will be revealed.

\end{document}